\documentclass[referee ]{sn-jnl}

\usepackage{graphicx}%
\usepackage{multirow}%
\usepackage{float}
\usepackage{amsmath,amssymb,amsfonts,bm}%
\usepackage{amsthm}%
\usepackage{mathrsfs}%
\usepackage[title]{appendix}%
\usepackage{xcolor}%
\usepackage[T1]{fontenc}
\usepackage{multirow}
\usepackage{float}
\usepackage{subcaption}
\usepackage{textcomp}%
\usepackage{manyfoot}%
\usepackage{booktabs}%
\usepackage{algorithm}%
\usepackage{algorithmicx}%
\usepackage{algpseudocode}%
\usepackage{listings}%
\usepackage{comment}
\usepackage{tabularx}
\usepackage[normalem]{ulem}
\begin{document}

\title[Functional forms]{Functional forms in joint models for longitudinal and time-to-event data: A practical guide with application and interpretation}


\author*[1,2]{\fnm{Felix Boakye} \sur{Oppong}}\email{felix.oppong@eortc.org}

\author[1]{\fnm{Dimitris} \sur{Rizopoulos}}

\author[2]{\fnm{Thierry} \sur{Gorlia}}

\author[3]{\fnm{Nicole} \sur{Erler}}

\affil[1]{\orgdiv{Department of Biostatistics, Erasmus University Medical Center,  \country{The Netherlands}}}

\affil[2]{\orgdiv{European Organisation for Research and Treatment of Cancer, \country{Belgium}}}

\affil[3]{\orgdiv{Julius Center for Health Sciences and Primary Care}, \orgname{University Medical Center Utrecht}, \country{The Netherlands}}


\abstract{
\textbf{Background:} Joint models for longitudinal and time‑to‑event data are widely used in clinical research, yet many of their most powerful features remain underutilised. In particular, the choice of functional form (association structure), which defines how the biomarker trajectory relates to event risk, is often treated as a technical detail despite its central role in shaping model assumptions and interpretation. As a result, researchers frequently rely on default specifications that may not capture clinically relevant characteristics of biomarker trajectories, potentially leading to misleading scientific conclusions.\\\\
\noindent
\textbf{Methods:} We provide a structured overview and practical guide to functional forms that link longitudinal and survival processes in joint models. We classify and compare commonly used and recently proposed association structures, including instantaneous effects (current value, slope, and acceleration), cumulative and change-based formulations, shared random effects, and variability-based associations. Using longitudinal white blood cell measurements and overall survival data from the MIRAGE glioblastoma trial, we illustrate how different functional forms target distinct features of the biomarker trajectory and define different biomarker–risk relationships.\\\\
\noindent
\textbf{Results:} Instantaneous functional forms capture the biomarker’s current level or short‑term dynamics, whereas cumulative and change‑based formulations reflect longer‑term exposure or trends over relevant past intervals. Variability‑based structures quantify instability in the biomarker trajectory, offering an alternative prognostic signal. As the interpretation of the association parameters depends on the chosen formulation, it's (implied) transformation and scale of the biomarker, and the time-scale, effect sizes are not directly comparable across functional forms, but the variety of functional forms can capture many of the complex relationships between outcomes present in clinical data. 
In the MIRAGE application, alternative functional forms produced different effect interpretations and, in some cases, different substantive conclusions regarding the biomarker–risk relationship.\\\\
\noindent
\textbf{Conclusions:} 
The choice of functional form is a key modelling decision in joint models and directly determines the interpretation of the biomarker–risk association. Careful alignment between the scientific question and the chosen functional form is essential for valid interpretation and transparent reporting in applied studies.
}




\maketitle

\section{Introduction}\label{sec1}

Joint models for longitudinal and time‑to‑event data have become essential for analysing how repeated biomarker measurements relate to a time‑to‑event endpoint \cite{Tsiatis2004JointOverview,Henderson2000JointData, Rizopoulos2012JointR}. These models offer a unified approach that jointly accounts for the longitudinal biomarker trajectory and the risk process governing event occurrence, thereby allowing us to more accurately capture the complex relationships between the biomarker dynamics and the underlying biological processes driving the event risk \cite{Rizopoulos2011ATime-to-event, Rizopoulos2014CombiningAveraging}. They play a central role in biomedical research, for example in oncology, cardiology, and HIV research, where biomarkers such as tumour burden, cardiac function, or immune cell counts evolve over time and directly influence prognosis \cite{Sartor1997RateRadiotherapy, Brown2009AssessingHIV/AIDS, McHunu2020JointTherapy, Joolharzadeh2023RecentCardio-Oncology}. \\\\
\noindent
In their simplest form, the event risk at any given time is assumed to depend either on the expected underlying value of the longitudinal biomarker or on subject‑specific random effects \cite{Rizopoulos2014CombiningAveraging, Ye2008SemiparametricApproach, Taylor2013Real-TimeModels}. While these formulations implicitly assume that instantaneous biomarker levels or latent subject‑specific deviations fully capture the relevant biological signal, the relationship between biomarker trajectories and event processes is often more complex. Different characteristics of a patient's biomarker history, such as the rate of change, cumulative exposure, or overall variability may influence the patient's prognosis in distinct ways \cite{Brown2009AssessingHIV/AIDS, Martins2022AHeterogeneity, Palma2025AVariability}. Ignoring such aspects may lead to incomplete or even misleading inference about the mechanisms linking biomarkers and clinical outcomes.\\\\
\noindent
To address this, several functional forms (also referred to as association structures) have been developed, each capturing a different aspect of the longitudinal process that may influence the hazard of an event. Among the instantaneous‑risk formulations, the time‑dependent slope association extends the current value structure by linking the hazard not only to the biomarker level but also to its instantaneous rate of change, thereby allowing the risk to depend on how the biomarker evolves over time rather than solely on its current level \cite{Ye2008SemiparametricApproach, Taylor2013Real-TimeModels, Yu2008IndividualModel, Brown2009AssessingHIV/AIDS}. Further extensions incorporate higher‑order features such as curvature, allowing the hazard to depend on the acceleration of the biomarker trajectory. Beyond these instantaneous effects, cumulative functional forms summarize the historical burden of the biomarker over time, either through area‑based measures that reflect overall exposure or through change‑based formulations that quantify deviations from earlier values \cite{Rizopoulos2012JointR, Rizopoulos2014CombiningAveraging, Brown2009AssessingHIV/AIDS}. Such cumulative effects may be particularly relevant when prolonged exposure influences risk, as seen in settings such as smoking exposure, radiation dose accumulation, etc. More recent developments introduce functional forms that account for subject‑specific variability or instability in the biomarker trajectory, capturing fluctuations that may themselves carry prognostic information \cite{Martins2022AHeterogeneity,Courcoul2025AEvents, Palma2025AVariability}. \\\
\noindent
The choice of functional form has important implications for model interpretation, estimation, and clinical inference. Each form embodies a different assumption about how the longitudinal process influences risk, whether through its current latent value, velocity, accumulated exposure, variability, etc. Therefore, understanding and appropriately selecting the functional form is crucial for accurate modeling and meaningful interpretation of joint model parameters.
\\\\
\noindent
In this paper, we provide a practical overview of the most commonly used functional forms in joint models for longitudinal and time-to-event data, and a novel formulation that links the biomarker variability to the event risk. Using longitudinal white blood cell count (WBC)  and overall survival data from the MIRAGE glioblastoma trial \cite{Roth2024MarizomibTrial}, we present illustrative examples for each functional form to demonstrate the implied relationships between the longitudinal trajectory and event risk. In addition,  extensions that allow more flexibility in modelling, including interactions with covariates and time-varying effects are also discussed.
\\\\
\noindent
The remainder of this paper is structured as follows. Section \ref{standard JM} describes the general formulation of a joint model for longitudinal and time-to-event data, and describes the example dataset. Section \ref{functional} presents the different functional forms, illustrates their application in practice, and provides guidance on the interpretation of the estimated parameters. Section \ref{discussion} discusses key considerations in choosing appropriate association structures and concludes with practical guidance on the interpretation, implementation, and selection of functional forms in joint models.
\section{Standard Joint Model and Real Data Example}\label{standard JM}
\begin{figure}[H]
    \centering
    \includegraphics[width=0.9\linewidth]{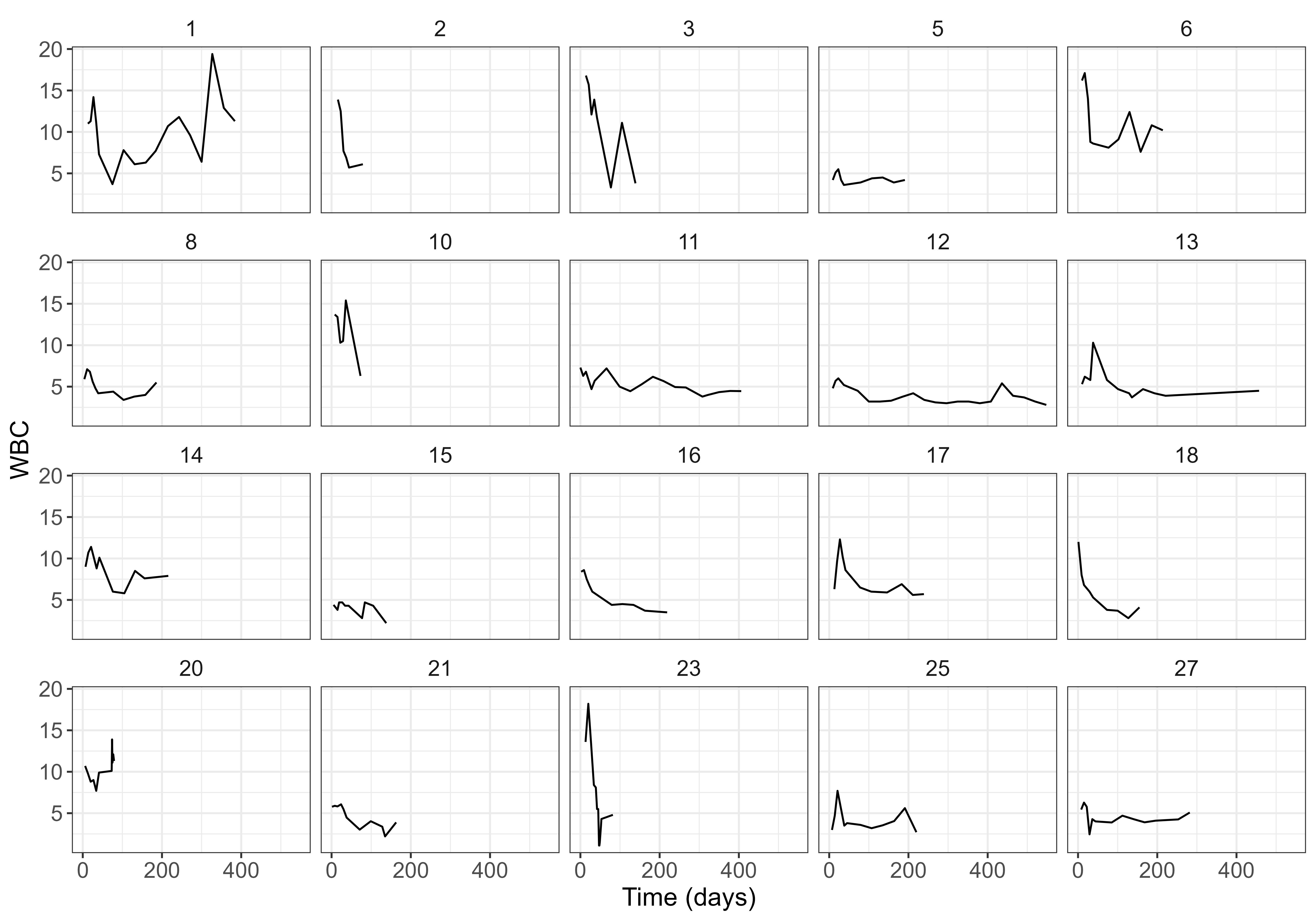}
    \caption{WBC count over time for the first 20 subjects}
    \label{fig:wbc}
\end{figure}

Joint models for longitudinal and survival data simultaneously model the evolution of time-dependent biomarkers and the occurrence of a time-to-event outcome \cite{Rizopoulos2012JointR, Tsiatis2004JointOverview}. For a continuous biomarker, the longitudinal submodel captures each subject's biomarker trajectory over time, typically using a linear mixed effects model:\\
\begin{equation}
\begin{aligned}
y_i(t) &= m_i(t) + \varepsilon_i(t), \\
m_i(t) &= \bm{x}_i^\top(t)\bm{\beta} + \bm{z}_i^\top(t)\bm{b}_i, \\
\bm{b}_i &\sim \mathcal{N}(\bm{0}, \bm{\Sigma}), \quad 
\varepsilon_i(t) \sim \mathcal{N}(0, \sigma^2),
\end{aligned}
\label{eq:model1}
\end{equation}
where $y_i(t)$ is the observed biomarker value for subject $i$ at time $t$, $m_i(t)$ is the true underlying biomarker value, $\bm{x}_i(t)$ and $\bm{z}_i(t)$ are design vectors for fixed and random effects $\bm{\beta}$ and $\bm{b}_i$, respectively, $\bm{\Sigma}$ is the covariance of the random effects, and $\varepsilon_i(t)$ is the error term. Beyond the linear mixed model, generalized linear models could also be considered for the longitudinal submodel.  \\\\
\noindent
The survival submodel links the longitudinal process to the hazard of the event using a proportional hazards model:

\begin{equation}
h_i\{t \mid \mathcal{M}_i(t), \bm{w}_i(t)\} = h_0(t) \exp\Big[ \bm{w}^\top_i(t)\bm{\gamma} + f\{t, \bm{b}_i, \mathcal{M}_i(t)\}^\top \bm{\alpha} \Big],
\label{eq:model2}
\end{equation}
\noindent 
where $\mathcal{M}_i(t) = \{m_i(u); 0 \leq u < t\}$ denotes the history of the true unobserved longitudinal process up to time $t$, $h_0(t)$ is the baseline hazard, and $\bm{w}_i(t)$ denotes baseline and time-dependent exogenous covariates with associated coefficients $\boldsymbol{\gamma}$. The association between the longitudinal biomarker process and the hazard is specified through the function $f\{t, \bm{b}_i, \mathcal{M}_i(t)\}$, with $\bm{\alpha}$ representing the corresponding vector of association parameters. \\\\
\noindent
Depending on the characteristic(s)  of the longitudinal biomarker believed to be related to the event risk, the function $f\{t, \bm{b}_i, \mathcal{M}_i(t)\}$ can take different forms, further detailed in the following Section.\\\\
\noindent
To illustrate the practical implementation and interpretation of these functional forms, data from the EORTC 1709 / Canadian Cancer Trials Group CE.8 (MIRAGE) trial \cite{Roth2024MarizomibTrial}, a multicentre, randomized phase 3 study in patients with newly diagnosed glioblastoma was used. This trial randomized a total of 749 patients to receive standard temozolomide-based radiochemotherapy with or without the experimental agent marizomib. Hematologic parameters, including white blood cell count (WBC) , were assessed weekly during the 6-week radiotherapy period, prior to each maintenance cycle, and at the end of treatment. For our illustration, we focus on WBC (unit: \(10^9\)/L) as a representative biomarker of immune function, which has previously been linked to survival outcomes in glioblastoma \cite{Gursoy2025CRP/albuminMultiforme, Zhang2025PeripheralTemozolomide}. Among the 749 randomized patients, 689 provided consent for re-use of their data for further research. After excluding 9 patients who had no data on WBC, 680 patients were included in these illustrative analyses. The median number of repeated measurements was 10, with a range of 1-43 measurements. This application focuses on illustration and interpretation, rather than model selection or formal comparison, providing guidance on specifying and understanding different functional forms in practice. The R package \textbf{JMbayes2} version 0.6-0 \cite{JMbayes2}  was used to fit the different joint models. In \textbf{JMbayes2}, the joint model is specified through separate fitted objects for the longitudinal and survival submodels, typically obtained from a linear mixed effects model and a Cox proportional hazards model, respectively.\\
\noindent
Given that the WBC measurements  exhibit non-linear trends over time (Figure~\ref{fig:wbc}), the longitudinal submodel was specified as a linear mixed effects model with natural cubic splines for time, including random effects for each subject. The corresponding linear mixed model can be fitted using the R package  \textbf{nlme} \cite{Pinheiro2019Nlme:Https://cran.r-project.org/package=nlme} with the following syntax:
\begin{verbatim}
fit_mixed_WBC <- lme(WBC ~ ns(time, 2), random = ~ ns(time, 2) | patid, 
            data = lab_data_WBC, control = lmeControl(opt = 'optim'))
\end{verbatim}
The survival submodel considered a Cox proportional hazards model for overall survival (OS), adjusted for treatment effect and other baseline covariates:
\begin{verbatim}
fit_cox_WBC <- coxph(Surv(tss, ss) ~ AGE + trt2 + strasurg + strakps 
               + nmgmt, data = pat_rand_WBC)
\end{verbatim}
Note that this proportional hazards model is only used to specify the exogenous part of the linear predictor and time-to-event outcome, and that the association between the (endogenous) longitudinal outome(s) and the event is specified separately, via the association structure.  How this ist done, is illustrated in the next section.
\noindent
The motivating example presented in this paper is based on the MIRAGE trial data which is not publicly available. To provide an illustration of the syntax for the different functional forms implemented in \textbf{JMbayes2}, a tutorial using the publicly available Primary Biliary Cholangitis (PBC) dataset is available at \url{https://t.ly/VqQhv}. The corresponding R Markdown source code is available in the associated GitHub repository: \url{https://t.ly/rj9hV}.

\section{Functional forms for the association structure}
\label{functional}

\begin{figure}[H]
    \centering
    \includegraphics[width=0.9\linewidth]{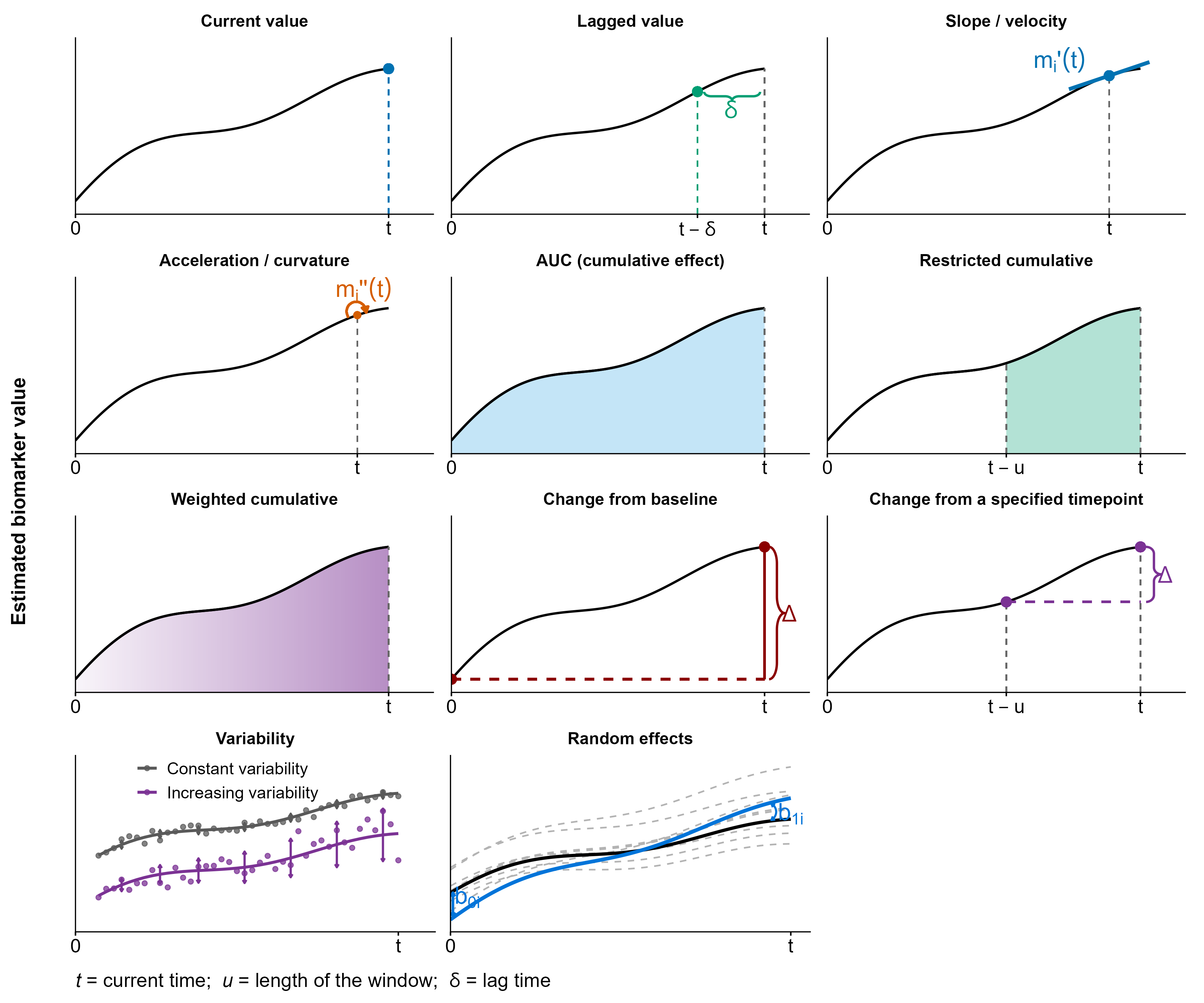}
    \caption{Graphical overview of the functional forms for linking longitudinal biomarkers to the hazard in joint models.}
    \label{fig:functional_forms}
\end{figure}

In the joint model described in Section \ref{standard JM}, the association between the longitudinal biomarker and the time-to-event outcome is fully determined by the choice of the function
\[
f\{t, \boldsymbol{b}_i, \mathcal{M}_i(t)\},
\]
which specifies how features of the subject-specific longitudinal process enter the hazard function. 
Although the survival submodel is formulated conditionally on the entire biomarker history
$\mathcal{M}_i(t) = \{m_i(u); 0 \le u < t\}$, the manner in which this history affects the hazard is determined by the specified association function
$f\{t, \boldsymbol{b}_i, \mathcal{M}_i(t)\}$. Hence, it is important to note that, including a longitudinal outcome in a joint model does not, by itself, imply that cumulative information from the biomarker history contributes to the event process. Rather, the relationship between the longitudinal and survival outcomes is governed by the chosen functional form, which determines which features of the biomarker trajectory are assumed to influence the hazard. Different specifications, therefore, encode different assumptions about the relevant aspects of the longitudinal process. Figure~\ref{fig:functional_forms} provides a graphical overview of the different functional forms presented in this paper. Each panel illustrates the feature of the estimated biomarker trajectory that is linked to the hazard under the corresponding association structure.\\\\
\noindent
In the following Sections, we distinguish between \textit{instantaneous  functional forms}, in which the risk of an event at time $t$ depends on instantaneous characteristics of the estimated biomarker trajectory evaluated at time $t$, and \textit{cumulative functional forms}, which summarise information from this trajetory over a (past) time interval. For each functional form, we provide a formal definition, describe its implementation in \textbf{JMbayes2}, and discuss the interpretation of the corresponding association parameters.

\subsection{Instantaneous association structures}
\label{current}

Instantaneous association structures link the instantaneous event risk at time $t$ to features of the longitudinal biomarker trajectory evaluated at the same time point. These specifications assume that the hazard depends on the current state or behaviour of the biomarker, rather than on its accumulated history. As such, they represent \textit{instantaneous} association structures, even though they are derived from an underlying longitudinal process.
\subsubsection{Current value functional form}\label{current-value}

\begin{figure}[H]
\centering
\includegraphics[width=1\textwidth]{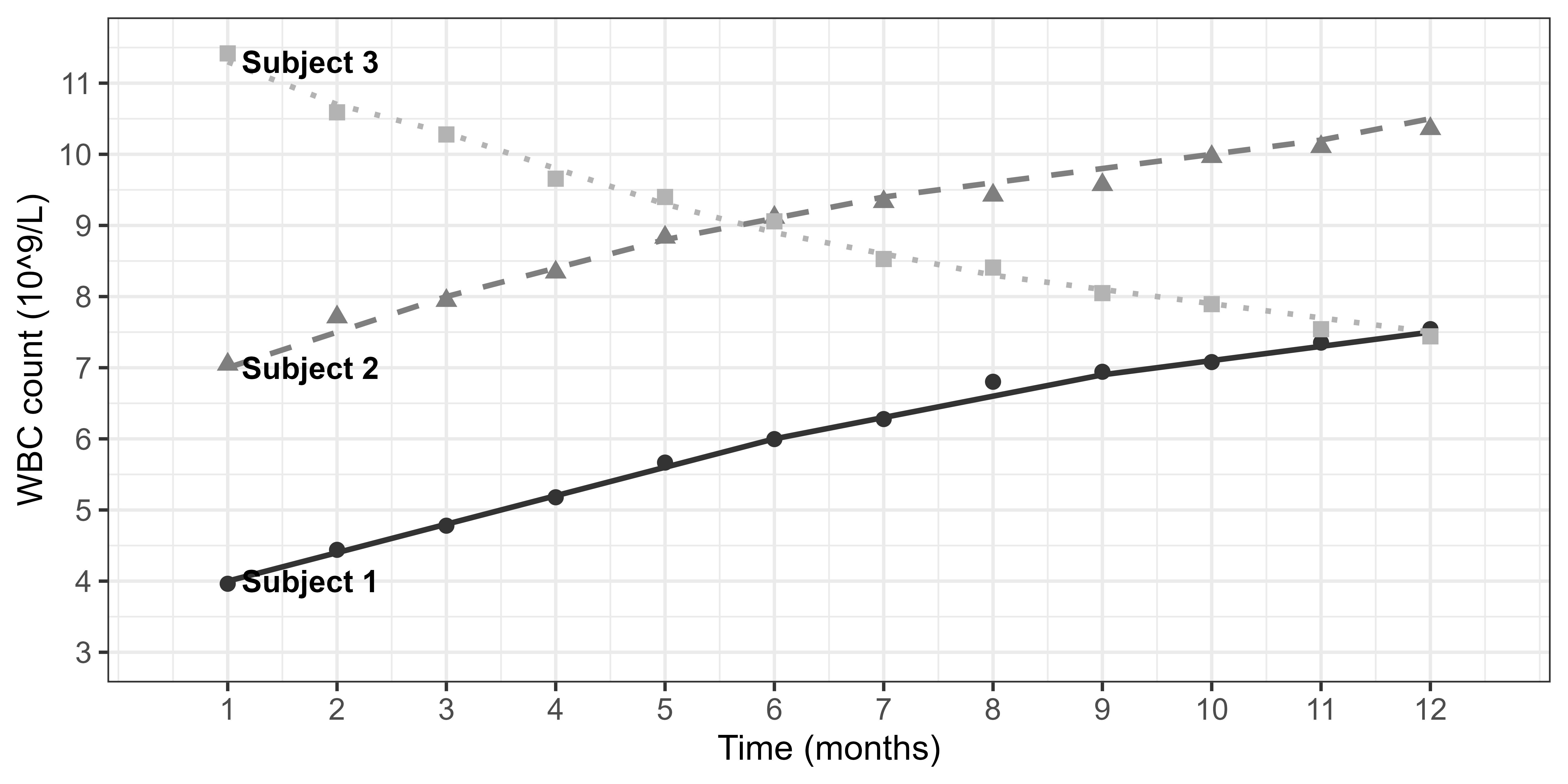}
\caption{ Illustration of functional forms using hypothetical data on WBC trajectory for 3 subjects over 12 months. The solid points represent observed values for each subject, while the lines show the estimated trajectories from a mixed model.
}
\label{fig:currentvalue}
\end{figure}

The simplest and most widely used association structure is based on the current value of the underlying longitudinal biomarker process. In this case, the association function is defined as
\begin{equation}
\begin{aligned}
f\{t, \boldsymbol{b}_i, \mathcal{M}_i(t)\} &= m_i(t),\\
&= \bm{x_i^\top}(t)\bm{\beta} + \bm{z_i^\top}(t)\bm{b_i},
\end{aligned}
\end{equation}
\noindent
where $m_i(t)$ denotes the subject-specific true underlying biomarker value at time $t$, as defined by the longitudinal submodel in~\eqref{eq:model1}. This functional form assumes that the instantaneous risk of event depends on the underlying current biomarker level. Although the survival submodel in~\eqref{eq:model2} is conditioned on the full biomarker history $\mathcal{M}_i(t)$, the current-value functional form reduces to $m_i(t)$, implying that distinct longitudinal trajectories may yield identical hazard contributions if their estimated values at time $t$ are the same. Consider the (artificial) example in Figure ~\ref{fig:currentvalue}. The three subjects have very different trajectories, however, the biomarker contribution to the risk at time 5.7 would be the same for subjects 2 and 3, and it would be the same for subjects 1 and 3 at time 12. The corresponding association parameter $\alpha$ quantifies the difference in the log-hazard for a one-unit difference in the biomarker value, holding all other covariates constant. Equivalently, $\exp(\alpha)$ represents the hazard ratio comparing two individuals who differ by one unit in their biomarker value at time $t$. 
\\\\
\noindent
For illustration, we fit a joint model for white blood cell (WBC) count and overall survival (OS) from the MIRAGE trial described in Section~\ref{standard JM}. Using \textbf{JMbayes2}, the joint model with the current value functional form can be fitted using the following syntax, where \texttt{fit\_cox\_WBC} and \texttt{fit\_mixed\_WBC} are the survival and longitudinal sub-models as specified above, \texttt{n\_chains} indicates the number of MCMC chains, and \texttt{n\_iter} and \texttt{n\_burnin} are the total number of iterations and number of burn-in iterations that will be discarded.  Since the current value functional form is the default in \textbf{JMbayes2}, the specification of the functional form via the argument \texttt{functional\_form = value(wbc)} is implied here.
\begin{verbatim}
jm_current_WBC <- jm(fit_cox_WBC, fit_mixed_WBC, time_var = "time", 
                  n_chains = 5L, n_iter = 10000L, n_burnin = 1000L)
\end{verbatim}
\begin{table}[h]
\centering
\caption{Posterior means and 95\% credible intervals from the survival submodel of the joint model assessing the association of white blood cell count (WBC) with overall survival using the current-value functional form.}
\label{tab:wbc_survival}
\begin{tabular}{lccc}
\hline
Covariate & Posterior Mean & 2.5\% CI & 97.5\% CI \\
\hline
Age (years) & 0.025 & 0.012 & 0.038 \\
Treatment (B vs.\ A) & 0.028 & -0.158 & 0.215 \\
Extent of Surgery (Biopsy/Partial vs.\ Gross total) & 0.367 & 0.197 & 0.535 \\
KPS (90/100 vs.\ 70/80) & -0.297 & -0.475 & -0.116 \\
MGMT methylated vs.\ unmethylated & -1.158 & -1.376 & -0.948 \\
MGMT undetermined/invalid vs.\ unmethylated & -0.300 & -0.611 & 0.000 \\
\textbf{Current WBC ($\alpha$)} & \textbf{0.011} & \textbf{-0.002} & \textbf{0.024}\\
\hline
\end{tabular}
\end{table}
\noindent
The model estimated that a $1\times10^9$/L higher WBC at any given time is associated with an approximate $1.1\%$ increase in the instantaneous log-hazard of death  (last row in table \ref{tab:wbc_survival}), suggesting a possible positive association, although with considerable uncertainty in this dataset as the 95\% credible interval includes zero. Table 1 also shows the results for the other covariates in the survival submodel: older age and Biopsy/Partial resection  are associated with higher risk, whereas better performance status and favorable molecular markers are protective. 
\\\\
\noindent
In some settings, it may be biologically implausible for the event risk at time $t$ to depend on the biomarker value observed at the exact same time point. Instead, a delayed effect may be more realistic. This can be accommodated by defining a lagged current value functional form \cite{Albert2010AnData}:
\begin{equation}
f\{t, \boldsymbol{b}_i, \mathcal{M}_i(t)\} = m_i(t - \delta), \qquad \delta > 0,
\end{equation}
where $\delta$ denotes a pre-specified lag. This specification assumes that the instantaneous hazard at time $t$ depends on the biomarker level measured $\delta$ time units earlier. Lagged current-value effects are particularly relevant in settings where the biological effect of a process is delayed. For example, following chemotherapy, bone marrow suppression develops over several days, such that the current risk of febrile neutropenia or infection may be more strongly related to the neutrophil count measured a few days earlier than to the contemporaneous value. In such situations, a lagged current value may better capture the biologically relevant state preceding the event.
\\\\
\noindent
An attractive feature of the (lagged) current value functional form is that the association parameter $\alpha$ is readily interpretable. As in a standard regression model, it represents the change in the log-hazard associated with a one-unit increase in the biomarker, evaluated either at the current time or $\delta$ time units earlier. However, it only makes limited use of the subject's biomarker history, and does not allow to distinguish subjects with very different biomarker trajectories, such as subjects 1 and 3 in Figure~\ref{fig:currentvalue}. For both, the biomarker would have the same contribution to the instantaneous risk at month 12 under the current value formulation, even though their longitudinal WBC trajectory is very different. 

\subsubsection{Time-dependent slope (velocity) functional form}\label{velocity}

The current value formulation presented in \ref{current-value}  does not encode if the biomarker is currently improving or deteriorating. The time-dependent slope functional form characterizes how the rate of change of a biomarker at time $t$ is associated with the risk of the event \cite{Andrinopoulou2017CombinedData, Rizopoulos2011ATime-to-event}. This association function is defined as
\begin{equation}
f\{t, \boldsymbol{b}_i, \mathcal{M}_i(t)\} = m_i'(t)  = \frac{d}{dt} m_i(t),
\end{equation}
that is, the first derivative of the estimated subject-specific longitudinal trajectory with respect to time. Here, the association parameter $\alpha$ quantifies the change in the log-hazard associated with a one-unit increase in the current rate of change of the biomarker per unit time. Intuitively, $m_i'(t)$ captures the instantaneous velocity of the biomarker, where positive values indicate increasing biomarker values, and negative values indicate a decreasing biomarker. 
\\\\
\noindent
Although closed-form expressions for the derivative $\frac{d}{dt} m_i(t)$
exist for many choices of the (non-linear) effect of time (i.e., the shape of the biomarker trajectories), these derivatives are not part of the mixed-effects model specification or estimation. They are only required when defining association structures in the survival submodel and must be derived explicitly when needed. This additional burden can be inconvenient in practice. Consequently, \textbf{JMbayes2} computes the velocity numerically using finite differences. Specifically, the derivative is approximated using central finite differences of the subject-specific fitted trajectory from the mixed-effects model. That is, if the biomarker is evaluated at time $t$, the velocity is computed as
\begin{equation}
\begin{aligned}
m_i'(t)
&\approx \frac{m_i(t+\epsilon) - m_i(t-\epsilon)}{2\epsilon} \\
&=
\tilde{\bm{x}}_i^\top(t)\bm{\beta}
+
\tilde{\bm{z}}_i^\top(t)\bm{b}_i,
\end{aligned}
\end{equation}
where $\epsilon$ is a small time increment (default $\epsilon = 0.001$ in \textbf{JMbayes2}), and the velocity design matrices are defined as
\[
\tilde{\bm{x}}_i^\top(t)
=
\frac{\bm{x}_i^\top(t+\epsilon) - \bm{x}_i^\top(t-\epsilon)}{2\epsilon},
\qquad
\tilde{\bm{z}}_i^\top(t)
=
\frac{\bm{z}_i^\top(t+\epsilon) - \bm{z}_i^\top(t-\epsilon)}{2\epsilon}.
\]
\noindent
This can evaluated at any time point for each subject, implicitly producing a latent time-varying covariate that enters the survival submodel. For a model with a linear effect of time, the time-dependent slope is constant over time. Therefore, the time-dependent slope functional form is most informative when the longitudinal trajectory includes nonlinear time effects, such as spline-based or polynomial terms, for which the instantaneous slope varies as a function of time.
\\\\
\noindent
Using the motivating dataset and the same longitudinal and survival submodels for WBC and OS described in Section~\ref{current-value}, the joint model using the time-dependent slope functional form was fitted by using the \texttt{update()} function to modify the previously specified model \texttt{(jm\_current\_WBC)} by changing only the specification of the association structure:

\begin{verbatim}
jm_velocity_WBC <- update(jm_current_WBC,
                   functional_forms = ~ velocity(WBC))
\end{verbatim}

\noindent
The estimated association parameter for the time-dependent slope for WBC is $\alpha = 1.402$, with a 95\% credible interval of  $[-1.821; 4.751]$. This suggests that subjects experiencing more rapid increases in WBC at a given time point tend to have a higher instantaneous risk of death compared with subjects whose WBC is stable or decreasing, even if their absolute WBC levels are similar. However, the wide credible interval includes zero, indicating substantial uncertainty in the estimated effect and weak evidence for a clear association between the instantaneous rate of change of WBC and overall survival in this dataset. \\\\
\noindent 
Importantly, interpretation of the time-dependent slope parameter may be less straightforward than that of the current-value functional form because the association is expressed on the scale of change in the biomarker per unit time rather than on the original biomarker scale. Consequently, the clinical meaning of a one-unit increase in the slope depends strongly on both the units of the biomarker and the time scale. For biomarkers that change gradually or are measured on a large scale, a one-unit increase in velocity may represent an unrealistically large change, making direct interpretation of the corresponding hazard ratio less intuitive. In this respect, the change (Delta) functional form, introduced in Section~\ref{change-general}, provides a related but often more interpretable alternative by quantifying changes in the biomarker over a clinically meaningful time interval rather than relying on an instantaneous rate of change.

\subsubsection{Acceleration functional form}\label{acceleration}

The acceleration functional form links the curvature of the longitudinal biomarker at time $t$ with the event risk. This functional form is defined as
\begin{equation}
f\{t, \boldsymbol{b}_i, \mathcal{M}_i(t)\} = m_i''(t) = \frac{d^2}{dt^2} m_i(t),
\end{equation}
where $m_i''(t)$ is the second derivative (with respect to time) of the expected value of the biomarker at time $t$. This formulation captures the changes in the velocity of the biomarker over time.

\noindent
As for the slope functional form, to unburden the user from having to specify the analytical solutions of the second derivative for the chosen function of time, \textbf{JMbayes2} computes the acceleration numerically using a second-order finite-difference approximation. Specifically, for a small time increment $\epsilon$, the acceleration at time $t$ is approximated as
\begin{equation}
\begin{aligned}
m_i''(t)
&\approx \frac{m_i(t+\epsilon) - 2 m_i(t) + m_i(t-\epsilon)}{\epsilon^2} \\
&=
\tilde{\tilde{\bm{x}}}_i^\top(t)\boldsymbol{\beta}
+
\tilde{\tilde{\bm{z}}}_i^\top(t)\bm{b}_i,
\end{aligned}
\end{equation}
where the acceleration design matrices are defined as
\[
\tilde{\tilde{\bm{x}}}_i^\top(t)
=
\frac{
\bm{x}_i^\top(t+\epsilon)
- 2 \bm{x}_i^\top(t)
+ \bm{x}_i^\top(t-\epsilon)
}{\epsilon^2},
\qquad
\tilde{\tilde{\bm{z}}}_i^\top(t)
=
\frac{
\bm{z}_i^\top(t+\epsilon)
- 2 \bm{z}_i^\top(t)
+ \bm{z}_i^\top(t-\epsilon)
}{\epsilon^2},
\]
\noindent
where $\epsilon$ is as previously defined. As with the velocity functional form, these quantities can be evaluated at any time point for each subject, implicitly producing a latent time-varying covariate that enters the survival submodel. For a model that assumes a linear effect of time, the acceleration equates to 0 (zero), which implies that acceleration is only meaningful when the longitudinal trajectory includes nonlinear time effect. \\\\
\noindent
To illustrate this functional form, we fitted a joint model for white blood cell (WBC) count and overall survival from the MIRAGE trial, using the same longitudinal and survival submodels described in Section~\ref{current-value}. The model was specified as follows:
\begin{verbatim}
jm_acceleration_WBC <- update(jm_current_WBC,  
                       functional_forms = ~ acceleration (WBC))
\end{verbatim}

\noindent
The estimated association parameter for the acceleration of WBC was $\alpha = 1464.33$ with a 95\% credible interval $[459.87, 2485.79]$, indicating that the curvature of the WBC trajectory is associated with overall survival. The positive estimate indicates that stronger upward curvature in the biomarker trajectory at time $t$ is associated with an increased instantaneous risk of death. Although the magnitude of the parameter estimate appears large, it reflects the scale of the acceleration term rather than an extreme biological effect. Because acceleration represents an instantaneous second‑order derivative of the biomarker trajectory, its numerical values are typically close to zero  (i.e., a change in acceleration of 1 would be extreme and likely biologically implausible), which naturally leads to larger regression coefficients in the survival submodel.

\subsection{Cumulative functional forms}
\label{cum}
For the instantaneous functional forms specification presented in Section~\ref{current}, the hazard at time $t$ is conditionally independent of the past biomarker history given the current value of the estimated trajectory, even though the longitudinal model itself is defined over time. For this, subjects with markedly different biomarker histories may contribute identically to the hazard at time $t$ if their estimated current biomarker values coincide. Cumulative functional forms address this limitation by linking the event risk at time $t$ to functions that capture parts of or the entire biomarker history up to that time \cite{Andrinopoulou2017CombinedData}. These specifications reflect the clinically plausible notion that prolonged exposure, rather than instantaneous biomarker values alone, may be relevant for the event process.

\subsubsection{Cumulative effect (area under the curve) functional form}
\label{cumulative}

The cumulative effect association structure links the hazard at time $t$ to the accumulated biomarker exposure from start to time $t$. The corresponding association function is

\begin{equation}
\begin{aligned}
f\{t, \boldsymbol{b}_i, \mathcal{M}_i(t)\}
&= \int_{0}^{t} m_i(s)\, ds \\
&= \int_{0}^{t}
\left\{
\bm{x}_i^\top(s)\boldsymbol{\beta}
+
\bm{z}_i^\top(s)\bm{b}_i
\right\}
ds .
\label{eq:model9}
\end{aligned}
\end{equation}
This formulation assumes that the entire biomarker history contributes to the hazard. As follow-up increases, however, the area under the fitted biomarker trajectory increases not only because of higher biomarker values but also because the integration interval becomes longer. This complicates the interpretation of the cumulative effect. To obtain a more interpretable measure, \textbf{JMbayes2} scales the integral by dividing it by $t$, yielding the average biomarker exposure over the observed follow-up period. Since the integral of the fitted biomarker trajectory generally does not have a closed-form solution, they are calculated using Gaussian quadrature, typically based on Gauss–Kronrod rules with adaptive refinement to control approximation error. Numerical quadrature approximates the integral by evaluating the trajectory at a set of carefully chosen time points (nodes) and combining these evaluations using corresponding weights that ensure high accuracy. Let $s_j^*$ and $w_j$ denote the quadrature nodes and weights mapped to the interval $[0,t]$. This functional form is approximated as follow:

\begin{equation}
\frac{1}{t}
\int_{0}^{t} m_i(s)\, ds
\approx 
\sum_{j=1}^{Q} w_j \, \bm{x}_i^\top(s_j^*) \boldsymbol{\beta}
+
  \sum_{j=1}^{Q} w_j \, \bm{z}_i^\top(s_j^*) \bm{b}_i\\
\end{equation}
\\
Using the example dataset with WBC count and OS, the joint model with the cumulative effect functional form was fitted in \textbf{JMbayes2} as follows:  

\begin{verbatim}
jm_area_WBC <- update(jm_current_WBC, functional_forms = ~ area(WBC))
\end{verbatim}

\begin{table}[h]
\centering
\caption{Posterior means and 95\% credible intervals for the survival submodel of the joint model using the cumulative (area under the trajectory) functional form for WBC.}
\label{tab:wbc_cumulative}
\begin{tabular}{lccc}
\hline
Covariate & Posterior Mean & 2.5\% CI & 97.5\% CI  \\
\hline
Age (years) & 0.025 & 0.013 & 0.038 \\
Treatment (B vs.\ A) & 0.024 & -0.164 & 0.209 \\
Extent of Surgery (Biopsy/Partial vs.\ Gross total) & 0.353 & 0.181 & 0.519 \\
KPS (90/100 vs.\ 70/80) & -0.274 & -0.454 & -0.090 \\
MGMT methylated vs.\ unmethylated & -1.147 & -1.367 & -0.936 \\
MGMT undetermined/invalid vs.\ unmethylated & -0.299 & -0.612 & -0.001 \\
Cumulative WBC (area) & 0.051 & 0.015 & 0.090 \\
\hline
\end{tabular}
\end{table}

\begin{figure}[H]
    \centering
\includegraphics[width=0.92\linewidth]{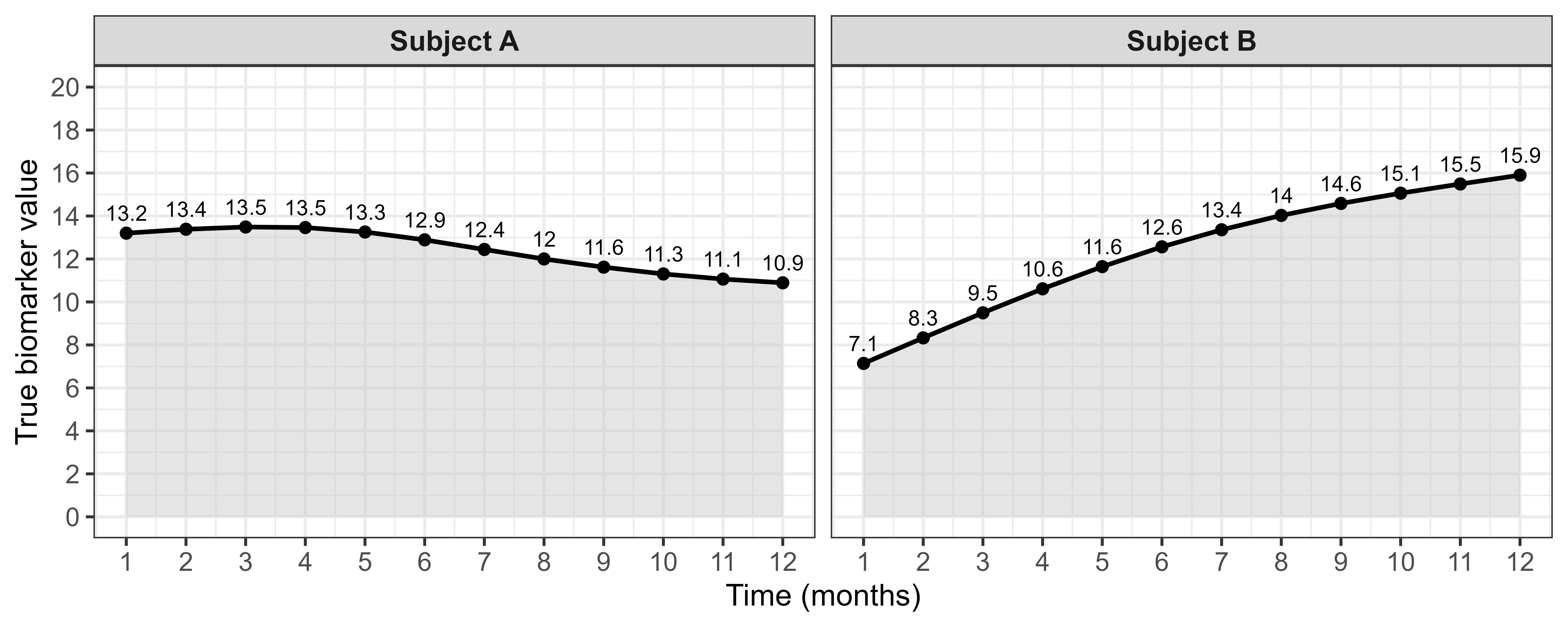}
    \caption{Two subjects with different estimated biomarker values at 12 months and different marker evolution}
    \label{fig:2}
\end{figure}

\noindent
Table~\ref{tab:wbc_cumulative} shows the posterior estimates from the survival submodel. The estimated association parameter for the cumulative WBC exposure is $\alpha = 0.051$ (95\% credible interval $[0.015, 0.090]$). This indicates that, at any given time, having a higher average past WBC over time is associated with a higher risk of death.
\\\\
\noindent
Even though the most common specification of the cumulative effect functional forms do explicitly take into account the entire history of the (estimated) biomarker profile, it is possible that distinct longitudinal trajectories may yield identical cumulative exposure values. For example, a patient with moderately elevated biomarker levels over a long period and another subject with very high levels over a short period may have the same area under the curve, despite very different temporal patterns. As illustrated in Figure \ref{fig:2}, although the two subjects have very different biomarker trajectories, their cumulative exposures at month 12 are equal, with  $\text{AUC} \approx 137$. 
\noindent
In some clinical settings, it may be reasonable to assume that the current risk of an event depends primarily on recent biomarker values rather than on the entire biomarker history. For example, short-term deterioration may be more clinically relevant than cumulative burden accumulated over several years. This is accommodated by the \textbf{restricted cumulative effect}, obtained directly from (\ref{eq:model9}) by choosing a fixed window length $u < t$, so that only biomarker values in the interval $(t - u, t]$ contribute to the hazard at time $t$. Returning to Figure~\ref{fig:2}, although Subjects A and B have similar total AUC over the 12-month period, the restricted cumulative effect distinguishes between their recent biomarker histories. It reflects differences in recent exposure between the two subjects, so that they are no longer assigned the same risk at $t = 12$.\\\\
\noindent
In \textbf{JMbayes2}, the restricted cumulative effect is specified via the \texttt{time\_window} argument. The example below links the hazard to the average WBC exposure over the preceding 30 days:
\begin{verbatim}
jm_area2_WBC <- update(jm_current_WBC, functional_forms = ~ 
                area(WBC, time_window = 30))
\end{verbatim}
\noindent
The estimated association parameter for the restricted cumulative WBC effect was $\alpha = 0.011$ with a 95\% credible interval $[-0.001, 0.024]$. This suggests that recent WBC exposure also contributes to the hazard of the event. Because the cumulative exposure measures are standardized by dividing the area under the trajectory by the length of the corresponding interval, the full and restricted cumulative effects are on the same scale. The smaller estimated association for the 30-day window therefore suggests that recent WBC history captures only part of the association between WBC and the hazard, whereas a measure incorporating the entire exposure history appears to be more strongly associated with the risk of death.

\subsubsection{Weighted cumulative functional form}
\label{weighted-cumulative}
In many biological processes, recent biomarker values may have a stronger influence on the risk of an event than older values. To allow different parts of the biomarker history to contribute unequally to the event risk, weighted cumulative functional forms extend the cumulative effect by introducing a weight function. The general form of the association function is

\begin{equation}
f\{t, \boldsymbol{b}_i, \mathcal{M}_i(t)\}
=
\int_{t-u}^{t} w(t - s)\, m_i(s)\, ds,
\end{equation}

\noindent
where $w(\cdot)$ is a nonnegative weight function that governs the relative importance of past biomarker values. Both the full cumulative effect and the restricted cumulative effect introduced in Section \ref{cumulative} can be viewed as special cases of the weighted cumulative functional form. Specifically, the full cumulative effect is recovered when $w(t-s)=1$ for all $s \in (t-u, t]$, assigning equal weight to all biomarker values within the exposure window. In contrast, the restricted cumulative effect is obtained by defining the weight function as

$$
w(t-s)=
\begin{cases}
1, & s \in (t-u, t],\\
0, & s \in [0, t-u].
\end{cases}
$$
More flexible formulations are also possible, in which a restricted time window is combined with a smoothly varying weight function taking values between 0 and 1, thereby allowing biomarker values within the window to contribute unequally to the hazard. Such weighted cumulative association structures have been proposed to flexibly characterize how the timing of biomarker exposure influences risk \cite{Mauff2017ExtensionEffects}. Although conceptually appealing, these extensions are currently not implemented in \textbf{JMbayes2}.

\subsubsection{Change ($\Delta$) functional forms}
\label{change-general}

Change-based association structures, sometimes referred to as \textit{Delta} ($\Delta$) functional forms, relate the hazard at time $t$ to the change in the biomarker up to time $t$. The corresponding association function is

\begin{equation}
\begin{aligned}
f\{t, \boldsymbol{b}_i, \mathcal{M}_i(t)\}
&= m_i(t)-m_i(0)\\
&=
\{\bm{x}_i(t)-\bm{x}_i(0)\}\boldsymbol{\beta}
+
\{\bm{z}_i(t)-\bm{z}_i(0)\}\bm{b}_i.
\end{aligned}
\end{equation}
This functional form represents the total change in the estimated biomarker level since baseline and is appropriate when deviation from the initial biomarker value is believed to be clinically meaningful. Since patients typically have different lengths of follow-up, it is often useful to standardize this change by dividing by the elapsed time $t$, thereby expressing the change per unit time. This standardized formulation is the default implementation in \textbf{JMbayes2}. The joint model with change in WBC count from baseline as the functional form was fitted as:

\begin{verbatim}
jm_delta_WBC <- update(jm_current_WBC, functional_forms = ~ Delta(WBC))
\end{verbatim}
The estimated association parameter is $\alpha = -4.355$
(95\% CI $[-13.271,4.443]$). This indicates that greater
deviation from baseline WBC may be associated with the
hazard of death, although the wide credible interval
indicates considerable uncertainty in this association. \\\\
\noindent
In some clinical settings, however, the current risk may depend more strongly on recent biomarker changes than on the total change since baseline. For example, a rapid deterioration during the preceding weeks may be more clinically relevant than gradual changes accumulated over a much longer period. This leads to the \textbf{restricted change} functional form, in which the change is evaluated over a fixed time window $(t-u,t]$, where $u>0$ denotes the length of the window. The association function becomes

\begin{equation}
\begin{aligned}
f\{t,\boldsymbol{b}_i,\mathcal{M}_i(t)\}
&=
m_i(t)-m_i(t-u)\\
&=
\{\bm{x}_i(t)-\bm{x}_i(t-u)\}\boldsymbol{\beta}
+
\{\bm{z}_i(t)-\bm{z}_i(t-u)\}\bm{b}_i.
\end{aligned}
\end{equation}
As for the change-from-baseline formulation, the change may be standardized by dividing by the window length $u$, yielding a rate of change per unit time. For our WBC example, we set the length of the window to 30 days (via the \texttt{time\_window} argument), to estimate the effect of change in WBC in the past 30 days as follows:

\begin{verbatim}
jm_delta30_WBC <- update(jm_current_WBC,
                   functional_forms = ~ Delta(WBC, time_window = 30))
\end{verbatim}

\noindent
The estimated association parameter was $\alpha=1.336$ (95\% credible interval $[-2.051, 4.843]$). This suggests that larger changes in WBC over a 30-day window may be associated with a higher instantaneous risk of death. However, the 95\% credible interval is wide and includes zero, indicating no clear evidence of an association in this dataset.
\\\\
\noindent
The change functional form is conceptually related to the time-dependent slope functional form described in Section~\ref{velocity}. Whereas the slope functional form captures the instantaneous rate of change at a specific time point through the derivative of the longitudinal trajectory, the change functional form summarizes change over a finite time interval. When the selected interval is short relative to the degree of nonlinearity in the biomarker trajectory, the standardized change may serve as a practical approximation to the instantaneous slope. In practice, the change functional form may also be easier to interpret clinically because it reflects biomarker changes over a familiar time window (e.g., 30 days), and the magnitude of change is often closer to what is observed in routine clinical monitoring.

\subsection{Shared random effects functional form}\label{random}

An association structure frequently used in the early joint modelling literature links the event risk directly to the subject-specific random effects from the longitudinal submodel \cite{Henderson2000JointData, Rizopoulos2012JointR, Wulfsohn1997AError}. In this formulation, the hazard at time $t$ depends only on the latent random effects that characterise individual deviations from the population-average biomarker trajectory. This functional form is defined as
\begin{equation}
\begin{aligned}
f\{t, \boldsymbol{b}_i, \mathcal{M}_i(t)\} &= \bm{b}_i,
\end{aligned}
\end{equation}
where $\boldsymbol{b}_i$ denotes the subject specific random effects from the longitudinal model. In this formulation, the association between the longitudinal process and the event risk is captured exclusively through the subject-specific random effects, rather than through explicit features of the estimated biomarker trajectory. While all functional forms discussed in this paper rely on random effects to construct the subject-specific trajectory, the shared random effects formulation links the hazard directly to these latent components themselves.

\noindent
A concrete illustration clarifies this distinction. Consider two patients with similar blood pressure trajectories over time, one older and one younger. Suppose age is included as a fixed effect in the longitudinal submodel, to account for the systematic increase in blood pressure associated with older age. After adjusting for this effect, the older patient may have a smaller random intercept than the younger patient, as part of their higher observed blood pressure is explained by age, whereas the younger patient requires a larger random intercept to reconcile their observed trajectory with the population-average trend. Random effects represent subject-specific deviations from the covariate-adjusted mean trajectory, and the shared random effects formulation links the hazard to these deviations. In contrast, functional forms based on the estimated trajectory relate the hazard to the estimated underlying biomarker level, which includes random as well as (fixed) covariate effects. \\\\
\noindent
Although incorporating such between-subject heterogeneity alongside other clinical covariates may be useful in certain prediction settings, the interpretation of the corresponding association parameter is less straightforward. This interpretational challenge becomes more pronounced in flexible longitudinal models, such as those based on penalised splines or other basis expansions, where the random effects correspond to basis coefficients rather than directly interpretable clinical quantities. In such cases, the subject-specific random effects no longer have an immediate clinical interpretation. Despite its historical importance and computational convenience \cite{Tsiatis2004JointOverview, Rizopoulos2012JointR}, the shared random effects formulation has attracted increasing criticism \cite{Rizopoulos2012JointR, Hickey2016JointIssues}, prompting a shift toward functional forms that link the hazard to interpretable aspects of the biomarker trajectory, such as its current value, its velocity, or cumulative exposure, among others.

\subsection{Variability-based functional form}\label{variability}

The functional forms introduced thus far are all defined through the general association function $f\{t, \boldsymbol{b}_i, \mathcal{M}_i(t)\},$
which links the hazard to features of the subject-specific mean trajectory $m_i(t)$ and the random effects $\boldsymbol{b}_i$. These functional forms are based on the estimated trajectory and do not account for the variability in the biomarker measurements  around this trajectory. As presented in Section \ref{standard JM} in the formulation of the standard joint model, it is assumed that the residual variability of the biomarker is constant across subjects and over time (see ~\ref{eq:model1}). This homoscedasticity assumption may be restrictive in practice, as individuals can differ not only in their mean trajectories but also in the variability of their biomarker measurements. Moreover, the short-term variability of biomarker measurements may itself carry prognostic information beyond the mean trajectory. For example, increased variability in clinical biomarkers has been associated with adverse outcomes in several disease areas, including oncology and cardiovascular disease, where greater instability in disease-related markers may reflect underlying physiological dysregulation and an increased risk of disease progression or mortality \cite{Piatek2020RisingCancer, Wu2023Visit-to-visitAnalysis}. To accommodate this  heteroscedasticity and model it's relationship with the event process, recent work has extended the standard formulation of the joint model to allow the residual variance to vary across subjects and time \cite{Palma2025AVariability, Courcoul2025AEvents, Li2023AOutcome}, leading to

\begin{equation}
\begin{aligned}
y_i(t) &= \bm{x}_i^\top(t)\bm{\beta} + \bm{z}_i^\top(t)\bm{b}_i + \varepsilon_i(t), \\
\bm{b}_i &\sim \mathcal{N}(\bm{0}, \bm{\Sigma}), \qquad 
\varepsilon_i(t) \sim \mathcal{N}(0, \sigma_i^2(t)),
\end{aligned}
\end{equation}
where $\sigma_i^2(t)$ is the subject-specific and time-dependent variance. To ensure positivity, $\sigma_i(t)$ is typically modeled on the log-scale via a separate submodel, for example
\[
\log \{\sigma_i(t)\} = \bm{u}_i^\top(t)\bm{\xi} + \bm{v}_i^\top(t)\bm{\eta}_i,
\]
with design vectors $\bm{u}_i(t)$ and $\bm{v}_i(t)$ and corresponding fixed and random effects. We define
$
\mathcal{V}_i(t) = \{ \sigma_i(u); 0 \le u < t \},
$
as the history of subject-specific standard deviations. The hazard can then be linked to features of this subject-specific standard deviation trajectory through the association function
$
f\{ t, \boldsymbol{\eta}_i, \mathcal{V}_i(t) \},
$
independently of the mean trajectory. The hazard model including the variability can therefore be written as
\begin{equation}
h_i(t \mid \mathcal{V}_i(t), \bm{w}_i(t)) = h_0(t) \exp \Big[
\bm{w}_i^\top(t) \bm{\gamma} + 
\bm{\alpha}_\sigma^\top f\{t, \boldsymbol{\eta}_i, \mathcal{V}_i(t)\} \Big].
\end{equation}

\noindent
The functional forms presented for the mean trajectory in Sections \ref{current} and  \ref{cum} can  analogously be applied to the variance trajectory. However, fitting a full heteroscedastic joint model requires dedicated software extensions, as standard joint modeling packages currently do not directly support subject-specific, time-varying residual variance. \\\\
\noindent
To overcome this, a variability-based functional form can be implemented using standard joint modeling software via a two-step approach \cite{Oppong2026JointApproach}. In the first step, residuals from a preliminary mixed-effects model are used as subject- and time-specific empirical approximations of within-subject variability. In the second step, a 
transformation of these residuals (e.g., the absolute residual $|\hat{\varepsilon}_i(t)|$ or the squared residual $\hat{\varepsilon}_i(t)^2$) is included as an additional 
longitudinal outcome in a standard joint model, with its fitted trajectory linked to the hazard through the association parameter $\alpha_\sigma$. Although this two-step approach does not fully account for the uncertainty associated with the estimated residual process and is therefore an approximation to the fully joint heteroscedastic model, simulation studies have shown that it can provide a reasonable approximation in practice. It may therefore serve as a pragmatic alternative when fitting a fully joint heteroscedastic model is computationally challenging or not feasible. This approach can readily be implemented in \textbf{JMbayes2} and other standard joint model software without any custom extensions.\\\\
\noindent
To illustrate this, we apply the two-step approach to model WBC count from the MIRAGE trial. Here, the longitudinal submodel is a linear mixed-effects model for the absolute residuals $|\hat{\varepsilon}_i(t)|$ as an approximation of the subject-specific standard 
deviation.
The model was fitted as follows:

\begin{verbatim}
# Step 1: fit preliminary mixed model for WBC and extract absolute residuals
fit_mixed_WBC <- lme(WBC ~  ns(time, 2), random = ~ ns(time, 2) | patid, 
              data = lab_data_WBC, control = lmeControl(opt = 'optim'))
lab_data_WBC$abs_resid <- abs(residuals(fit_mixed_WBC))

# Step 2: pre-fit mixed model for variability (input for joint model)
fit_mixed_var <- lme(abs_resid ~ time, random = ~ time | patid, 
               data = lab_data_WBC)
# The cox model input for the joint model is fit_cox_WBC, as specified 
in Section 2
# Joint model with WBC count variability
jm_variability_WBC <- jm(fit_cox_WBC, fit_mixed_var, time_var = "time",
                    n_chains = 5L, n_iter = 10000L, n_burnin = 1000L)
\end{verbatim}
\noindent
The estimated association parameter was $\hat{\alpha}_{\varepsilon}=0.236$ (95\% credible interval $[-0.045, 0.549]$). This positive estimate suggests that greater within-subject variability in WBC count may be associated with an increased hazard of death, although the wide credible interval indicates substantial uncertainty regarding the strength of this association. Clinically, this may reflect the idea that instability in biomarker measurements, beyond the mean trajectory itself, captures additional prognostic information. For example, two patients with similar average WBC trajectories may differ in risk if one exhibits substantially greater fluctuations over time. \\\\
\noindent
While the present example linked the hazard to the current level of biomarker variability, the two-step framework is sufficiently flexible to accommodate any of the functional forms described in Sections  \ref{current} and  \ref{cum}. For example, one could model the association between the hazard and the slope, cumulative exposure, or recent change in biomarker variability over time. However, although such extensions are theoretically possible, the resulting association parameters may not always admit intuitive clinical interpretations. For instance, relating the hazard to the velocity of the biomarker standard deviation may be mathematically well defined, but its clinical meaning may be difficult to motivate or communicate. Consequently, the choice of functional form for variability-based associations should be guided not only by statistical flexibility but also by interpretability and clinical plausibility. \\\\
\noindent
An overview of all functional forms discussed in this Section, including their definitions, interpretation, and the aspect of the biomarker trajectory they capture, is presented in table \ref{tab:functional_forms_summary}.

\begin{table}[!h]
\centering
\caption{Summary of functional forms for linking longitudinal biomarkers to the hazard in joint models}
\label{tab:functional_forms_summary}
\renewcommand{\arraystretch}{1.2}
\begin{tabularx}{\textwidth}{X X X X}
\hline
\textbf{Functional form} & \textbf{Quantity} & \textbf{Scale / units} & \textbf{Interpretation of $\alpha$} \\
\hline

Current value & Estimated biomarker level at time $t$ & Original biomarker scale & Change in log-hazard per unit difference in current biomarker level \\
\hline
Time-dependent slope & Instantaneous rate of change & Biomarker per unit time & Change in log-hazard per unit difference in velocity of biomarker (i.e., change in biomarker per time unit) \\\hline

Acceleration & Curvature (change in slope) & Biomarker per time$^2$ & Change in log-hazard per unit increase in velocity.. Highly scale-sensitive association \\
\hline

Change (Delta) & Change over a specified window $u$ & Biomarker (or per time unit if standardized) & Change in log-hazard per unit difference in biomarker change over the specified window \\\hline

Cumulative (AUC) & Total historical exposure & Biomarker $\times$ time & Change in log-hazard per unit difference in accumulated biomarker exposure up to time $t$\\
\hline

Standardized cumulative (mean exposure)  & Average historical exposure & Original biomarker scale & Change in log-hazard per unit difference in average biomarker exposure up to time $t$\\
\hline
Restricted cumulative & Recent exposure over window $u$ & Biomarker $\times$ time (window-based) & Change in log-hazard per unit difference in accumulated biomarker exposure over the specified window \\
\hline

Shared random effects & Latent subject-specific deviations & Units depend on the associated random effect (e.g. biomarker units for a random intercept; biomarker per unit time for a random slope) & Change in log-hazard per unit difference in the corresponding random effect(s) \\
\hline
Variability-based & Within-subject variability over time & Biomarker variability scale (e.g. SD or variance) & Change in log-hazard per unit difference in biomarker variability \\
\hline
\end{tabularx}
\end{table}

\section{Discussion}\label{discussion}

In this paper, we provided a practical overview of functional forms for linking longitudinal biomarker trajectories to time-to-event outcomes in a joint model. We showed how each association structure captures a different aspect of the biomarker process and how the choice of functional form shapes both the interpretation of the estimated parameters and the resulting scientific conclusions. Using data from the MIRAGE trial, we demonstrated how these structures can be implemented in practice with the R package \textbf{JMbayes2}.\\

\noindent
The functional forms considered in this work can be broadly grouped according to how they use information from the biomarker trajectory. Instantaneous functional forms, including the current value, time-dependent slope, and acceleration, link the hazard at time $t$ to features of the biomarker trajectory evaluated at that same time point. Although the current-value formulation provides a natural and interpretable association parameter, it does not explicitly incorporate earlier biomarker values beyond their contribution to the current estimated biomarker level \cite{Rizopoulos2012JointR, Taylor2013Real-TimeModels}. Consequently, subjects with identical estimated biomarker values at a given time may have similar contributions to the hazard even if their earlier biomarker trajectories differ substantially. The slope and acceleration formulations extend this framework by incorporating the velocity and curvature of the trajectory \cite{Brown2009AssessingHIV/AIDS, Yu2008IndividualModel}, which may be informative when biomarker dynamics are biologically relevant.\\

\noindent
In contrast, cumulative functional forms summarize the biomarker history over time. These association structures reflect the idea that risk may depend on sustained exposure rather than on instantaneous levels alone \cite{Brown2009AssessingHIV/AIDS, Mauff2017ExtensionEffects}. They are particularly relevant when the underlying mechanism operates through long-term burden. Change-based formulations quantify the effect of change in a biomarker over a defined period \cite{Ye2008SemiparametricApproach}, which may be more relevant when recent deterioration or improvement is of interest. The variability-based functional form captures fluctuations around the mean trajectory and reflects the notion that instability in a biomarker may be a prognostic signal \cite{Courcoul2025AEvents, Palma2025AVariability, Martins2022AHeterogeneity}. Finally, the shared random effects formulation links the hazard to subject specific deviations from a population trajectory which can complicate interpretation and may obscure the underlying biological mechanism. Given the availability of flexible software such as \textbf{JMbayes2}, trajectory-based formulations are often preferable when the scientific objective is to understand how biomarker dynamics relate to risk.\\

\noindent
From a practical perspective, the choice of functional form should be guided by the hypothesized mechanism linking the biomarker to the event process. Different formulations may be appropriate depending on whether risk is thought to depend primarily on the current biomarker level, its recent evolution, cumulative exposure, variability, or other trajectory characteristics. When no clear biological rationale exists, comparing several clinically plausible formulations and evaluating their predictive performance may be informative.\\

\noindent
We also note that the scale of the association parameter depends fundamentally on the chosen functional form and the scale of the underlying covariate entering the survival model. Each functional form induces a different scale, which is central to the interpretation of the corresponding parameter. As a result, the magnitude of association parameters is not directly comparable across functional forms, and differences in scale should not be interpreted as differences in effect size. Instead, interpretation should be grounded in the clinical meaning of the underlying transformation of the biomarker trajectory. Meaningful clinical interpretation therefore requires understanding both the mathematical transformation and the time scale on which it is defined.\\

\noindent
When the longitudinal outcome is modelled on a transformed scale, for example using a log transformation to address skewness, the subject-specific linear predictor from the mixed-effects model does not lie on the natural scale of the biomarker. This can make the resulting association parameter difficult to interpret clinically because the functional form is evaluated on the transformed rather than the original scale. In such cases, it may be desirable to transform the subject-specific linear predictor back to the outcome scale before linking it to the hazard. Transformation functions provide a flexible mechanism to achieve this within \textbf{JMbayes2} \cite{JMbayes2}, allowing the association parameter to be interpreted on a clinically meaningful scale.\\

\noindent
Functional forms need not be considered in isolation and may be combined when supported by the scientific question \cite{Rizopoulos2014CombiningAveraging, Taylor2013Real-TimeModels}. For example, both the current biomarker level and its rate of change may jointly influence risk, or biomarker variability may provide prognostic information beyond the mean trajectory. Such combinations can capture complementary biological mechanisms but require careful consideration of identifiability, interpretability, and clinical plausibility.\\

\noindent
By default, the association between the longitudinal biomarker and the hazard is governed by a time-constant coefficient, consistent with the proportional hazards assumption in the survival submodel. In many clinical settings, however, the strength or even the direction of the association may evolve over time \cite{Sartor1997RateRadiotherapy}. For example, a biomarker may be strongly prognostic early in follow-up but less informative at later time points, or its prognostic value may depend on treatment phase or disease progression. To accommodate such scenarios, the joint model can be extended by allowing the association parameter to vary over time through an interaction between the functional form covariate and a function of time \cite{Lin2002MaximumVariables}. This extension relaxes the proportional hazards assumption for the longitudinal association, allowing the effect of the biomarker on the log-hazard to change over follow-up rather than remaining constant. Such flexibility can be applied to any of the functional forms introduced in this paper and is possible using software such as \textbf{JMbayes2}.\\

\noindent
Some limitations of the current overview should be acknowledged. First, while we have illustrated each functional form using a single dataset, the relative performance of different association structures will depend on the clinical context, the measurement frequency, and the underlying biological process. Formal model comparison may be useful when the scientific question does not clearly favour one functional form over another, but should not replace subject-matter reasoning. Second, the variability-based functional form as implemented via the two-step approach inherits the limitations of two-stage methods more generally. Uncertainty from the first-stage estimation is not propagated to the second stage, which may lead to underestimated standard errors for the association parameters \cite{Tsiatis2004JointOverview, Guler2014JointData}.\\

\noindent
Despite these limitations, this work has several strengths. We provide applied researchers with  a practical overview of a broad range of functional forms and illustrate their implementation and interpretation using a clinical trial dataset and \textbf{JMbayes2}. In addition, the inclusion of less commonly discussed formulations, such as curvature and variability-based functional forms, broadens the scope of available association structures and highlights opportunities for future methodological development.\\

\noindent
In conclusion, the choice of functional form in a joint model is not merely a technical modelling decision but a scientific one that determines how the longitudinal biomarker process is related to event risk \cite{Rizopoulos2012JointR, Hickey2016JointIssues, Brown2009AssessingHIV/AIDS, Andrinopoulou2017CombinedData}. No single association structure will be universally appropriate. Selection should therefore be guided by the clinical question, biological plausibility, interpretability, and the aspect of the biomarker trajectory believed to be most relevant to the outcome.


\section*{Declarations}

\begin{itemize}
\item Funding: This work is supported by the EORTC and the Belgian National Lottery and their players.
\item Competing interests: The authors declare no competing interests.
\item Ethics approval and consent to participate: Not applicable.
\item Consent for publication: Not applicable.
\item Data availability: The example dataset can be obtained from the EORTC through datasharing request via \url{https://www.eortc.org/data-sharing/}
\item Author contribution: F.B.O, D.R. and N.E. conceived the idea. D.R. and N.E. provided statistical guidance and supervision. F.B.O. conducted the methodological review, performed the statistical analyses and illustrative examples, developed the figures, and drafted the manuscript. T.G. provided clinical guidance and supervision, including input on the motivating example and interpretation of the clinical context. All authors contributed to the interpretation of the results, critically reviewed and revised the manuscript, and approved the final version.
\end{itemize}

\noindent

\bibliography{JM.bib, ref.bib}

@article{Palma2025AVariability,
    title = {{A Bayesian location-scale joint model for time-to-event and multivariate longitudinal data with association based on within-individual variability}},
    year = {2025},
    journal = {arXiv:2503.12270},
    author = {Palma, Marco and Keogh, Ruth H and Carr, Siobhán B and Szczesniak, Rhonda and Taylor-Robinson, David and Wood, Angela M and Muniz-Terrera, Graciela and Barrett, Jessica K},
    number = {},
    month = {3},
    url = {http://arxiv.org/abs/2503.12270},
    arxivId = {2503.12270}
}

@article{Rizopoulos2011ATime-to-event,
    title = {{A Bayesian semiparametric multivariate joint model for multiple longitudinal outcomes and a time-to-event}},
    year = {2011},
    journal = {Statistics in Medicine},
    author = {Rizopoulos, Dimitris and Ghosh, Pulak},
    number = {12},
    pages = {1366--1380},
    volume = {30},
    doi = {10.1002/sim.4205},
    issn = {02776715}
}

@article{Martins2022AHeterogeneity,
    title = {{A flexible link for joint modelling longitudinal and survival data accounting for individual longitudinal heterogeneity}},
    year = {2022},
    journal = {Statistical Methods and Applications},
    author = {Martins, Rui},
    number = {1},
    pages = {41–61},
    volume = {31},
    doi = {10.1007/s10260-021-00566-6},
    issn = {1613981X}
}

@article{Wulfsohn1997AError,
    title = {{A Joint Model for Survival and Longitudinal Data Measured with Error}},
    year = {1997},
    journal = {Biometrics},
    author = {Wulfsohn, Michael S. and Tsiatis, Anastasios A.},
    number = {1},
    pages = {330--9},
    volume = {53},
    doi = {10.2307/2533118},
    issn = {0006341X}
}

@article{Li2023AOutcome,
    title = {{A joint model of the individual mean and within-subject variability of a longitudinal outcome with a competing risks time-to-event outcome}},
    year = {2023},
    journal = {arXiv:2301.06584},
    author = {Li, Shanpeng and Nuyujukian, Daniel S. and McClelland, Robyn L. and Reaven, Peter D. and Zhou, Jin and Zhou, Hua and Li, Gang},
    number = {},
    pages = {},
    volume = {},
    url = {http://arxiv.org/abs/2301.06584},
    doi = {10.48550/arXiv.2301.06584},
    arxivId = {2301.06584}
}

@article{Courcoul2025AEvents,
    title = {{A Location-Scale Joint Model for Studying the Link Between the Time-Dependent Subject-Specific Variability of Blood Pressure and Competing Events}},
    year = {2025},
    journal = {Statistics in Medicine},
    author = {Courcoul, Léonie and Tzourio, Christophe and Woodward, Mark and Barbieri, Antoine and Jacqmin-Gadda, Hélène},
    number = {20-22},
    month = {9},
    volume = {44},
    publisher = {John Wiley and Sons Ltd},
    doi = {10.1002/sim.70244},
    issn = {10970258},
    pmid = {40911363},
    arxivId = {2306.16785}
}

@article{Albert2010AnData,
    title = {{An approach for jointly modeling multivariate longitudinal measurements and discrete time-to-event data}},
    year = {2010},
    journal = {Annals of Applied Statistics},
    author = {Albert, Paul S. and Shih, Joanna H.},
    number = {3},
    pages = {1517--1532},
    volume = {4},
    doi = {10.1214/10-AOAS339},
    issn = {19326157}
}

@article{Brown2009AssessingHIV/AIDS,
    title = {{Assessing the association between trends in a biomarker and risk of event with an application in pediatric HIV/AIDS}},
    year = {2009},
    journal = {Annals of Applied Statistics},
    author = {Brown, Elizabeth R.},
    number = {3},
    pages = {1163--1182},
    volume = {3},
    doi = {10.1214/09-AOAS251},
    issn = {19326157}
}

@article{Andrinopoulou2017CombinedData,
    title = {{Combined dynamic predictions using joint models of two longitudinal outcomes and competing risk data}},
    year = {2017},
    journal = {Statistical Methods in Medical Research},
    author = {Andrinopoulou, Eleni Rosalina and Rizopoulos, D. and Takkenberg, Johanna J.M. and Lesaffre, E.},
    number = {4},
    pages = {1787--1801},
    volume = {26},
    doi = {10.1177/0962280215588340},
    issn = {14770334}
}

@article{Rizopoulos2014CombiningAveraging,
    title = {{Combining Dynamic Predictions From Joint Models for Longitudinal and Time-to-Event Data Using Bayesian Model Averaging}},
    year = {2014},
    journal = {Journal of the American Statistical Association},
    author = {Rizopoulos, Dimitris and Hatfield, Laura A. and Carlin, Bradley P. and Takkenberg, Johanna J.M.},
    number = {508},
    pages = {1385–1397},
    volume = {109},
    doi = {10.1080/01621459.2014.931236},
    issn = {1537274X}
}

@article{Gursoy2025CRP/albuminMultiforme,
    title = {{CRP/albumin ratio and WBC values correlate with Ki-67 and survival in glioblastoma multiforme}},
    year = {2025},
    journal = {Frontiers in Oncology},
    author = {G{\"{u}}rsoy, Güven},
    pages = {1612212},
    volume = {15},
    publisher = {Frontiers Media SA},
    doi = {10.3389/fonc.2025.1612212},
    issn = {2234943X}
}

@article{Mauff2017ExtensionEffects,
    title = {{Extension of the association structure in joint models to include weighted cumulative effects}},
    year = {2017},
    journal = {Statistics in Medicine},
    author = {Mauff, Katya and Steyerberg, Ewout W. and Nijpels, Giel and van der Heijden, Amber A.W.A. and Rizopoulos, Dimitris},
    number = {23},
    pages = {3746--3759},
    volume = {36},
    doi = {10.1002/sim.7385},
    issn = {10970258}
}

@article{Yu2008IndividualModel,
    title = {{Individual prediction in prostate cancer studies using a joint longitudinal survival-cure model}},
    year = {2008},
    journal = {Journal of the American Statistical Association},
    author = {Yu, Menggang and Taylor, Jeremy M.G. and Sandler, Howard M.},
    number = {481},
    pages = {178--187},
    volume = {103},
    doi = {10.1198/016214507000000400},
    issn = {01621459}
}

@article{Tsiatis2004JointOverview,
    title = {{Joint modeling of longitudinal and time-to-event data: An overview}},
    year = {2004},
    journal = {Statistica Sinica},
    author = {Tsiatis, Anastasios A. and Davidian, Marie},
    number = {3},
    pages = {809--834},
    volume = {14},
    issn = {10170405}
}

@article{Guler2014JointData,
    title = {{Joint modelling for longitudinal and time-to-event data: Application to liver transplantation data}},
    year = {2014},
    journal = {Lecture Notes in Computer Science (including subseries Lecture Notes in Artificial Intelligence and Lecture Notes in Bioinformatics)},
    author = {Guler, Ipek and Calaza-D{\'{i}}az, Laura and Faes, Christel and Cadarso-Su{\'{a}}rez, Carmen and Giraldez, Elena and Gude, Francisco},
    number = {PART 3},
    pages = {580--593},
    volume = {8581 LNCS},
    isbn = {9783319091495},
    doi = {10.1007/978-3-319-09150-1{\_}42},
    issn = {16113349}
}

@article{McHunu2020JointTherapy,
    title = {{Joint modelling of longitudinal and time-to-event data: An illustration using CD4 count and mortality in a cohort of patients initiated on antiretroviral therapy}},
    year = {2020},
    journal = {BMC Infectious Diseases},
    author = {McHunu, Nobuhle N. and Mwambi, Henry G. and Reddy, Tarylee and Yende-Zuma, Nonhlanhla and Naidoo, Kogieleum},
    number = {1},
    pages = {256},
    volume = {20},
    doi = {10.1186/s12879-020-04962-3},
    issn = {14712334}
}

@article{Henderson2000JointData,
    title = {{Joint modelling of longitudinal measurements and event time data}},
    year = {2000},
    journal = {Biostatistics},
    author = {Henderson, R. and Diggle, P. and Dobson, A.},
    number = {4},
    pages = {465--480},
    volume = {1},
    doi = {10.1093/biostatistics/1.4.465},
    issn = {14654644}
}

@article{Oppong2026JointApproach,
    title = {{Joint modelling of time-dependent biomarker variability and time-to-event outcomes, a two-step approach}},
    year = {2026},
    journal = {arXiv:2605.05923},
    author = {Oppong, Felix Boakye and Rizopoulos, Dimitris and Gorlia, Thierry and Erler, Nicole},
    month = {5},
    url = {http://arxiv.org/abs/2605.05923},
    arxivId = {2605.05923}
}

@article{Hickey2016JointIssues,
    title = {{Joint modelling of time-to-event and multivariate longitudinal outcomes: Recent developments and issues}},
    year = {2016},
    journal = {BMC Medical Research Methodology},
    author = {Hickey, Graeme L. and Philipson, Pete and Jorgensen, Andrea and Kolamunnage-Dona, Ruwanthi},
    number = {1},
    pages = {117},
    volume = {16},
    doi = {10.1186/s12874-016-0212-5},
    issn = {14712288}
}

@book{Rizopoulos2012JointR,
    title = {{Joint Models for Longitudinal and Time-to-Event Data With Applications in R}},
    year = {2012},
    author = {Rizopoulos, Dimitris},
    publisher = {Chapman and Hall/CRC.},
    address = {Boca Raton},
    isbn = {9781447128977},
    doi = {10.1007/978-1-4471-2897-7}
}

@article{Roth2024MarizomibTrial,
    title = {{Marizomib for patients with newly diagnosed glioblastoma: A randomized phase 3 trial}},
    year = {2024},
    journal = {Neuro-Oncology},
    author = {Roth, Patrick and Gorlia, Thierry and Reijneveld, Jaap C. and de Vos, Filip and Idbaih, Ahmed and Frenel, Jean Sébastien and Rhun, Emilie Le and Sepulveda, Juan Manuel and Perry, James and Masucci, G. Laura and Freres, Pierre and Hirte, Hal and Seidel, Clemens and Walenkamp, Annemiek and Lukacova, Slavka and Meijnders, Paul and Blais, Andre and Ducray, Francois and Verschaeve, Vincent and Nicholas, Garth and Balana, Carmen and Bota, Daniela A. and Preusser, Matthias and Nuyens, Sarah and Dhermain, Fréderic and van den Bent, Martin and O’Callaghan, Chris J. and Vanlancker, Maureen and Mason, Warren and Weller, Michael},
    number = {9},
    pages = {1670--1682},
    volume = {26},
    doi = {10.1093/neuonc/noae053},
    issn = {15235866}
}

@article{Lin2002MaximumVariables,
    title = {{Maximum likelihood estimation in the joint analysis of time-to-event and multiple longitudinal variables}},
    year = {2002},
    journal = {Statistics in Medicine},
    author = {Lin, Haiqun and McCulloch, Charles E. and Mayne, Susan T.},
    number = {16},
    pages = {2369--82},
    volume = {21},
    doi = {10.1002/sim.1179},
    issn = {02776715}
}

@misc{Pinheiro2019Nlme:Https://cran.r-project.org/package=nlme,
    title = {{nlme: Linear and nonlinear mixed effects models. https://cran.r-project.org/package=nlme}},
    year = {2019},
    booktitle = {R-project},
    author = {Pinheiro, J and Bates, D and DebRoy, S and Sarkar, D and Team, R Core}
}

@article{Zhang2025PeripheralTemozolomide,
    title = {{Peripheral biomarkers predict survival in patients with glioblastoma treated with temozolomide}},
    year = {2025},
    journal = {Molecular and Clinical Oncology},
    author = {Zhang, Shuodan and McMillan, Nadia and McGuinness, Matthew and Trudeau, Stephen and Ho, Ka Wai Grace and Uhlmann, Erik J.},
    number = {6},
    pages = {56},
    volume = {22},
    publisher = {Spandidos Publications},
    doi = {10.3892/mco.2025.2851},
    issn = {20499469}
}

@article{Sartor1997RateRadiotherapy,
    title = {{Rate of PSA rise predicts metastatic versus local recurrence after definitive radiotherapy}},
    year = {1997},
    journal = {International Journal of Radiation Oncology Biology Physics},
    author = {Sartor, Carolyn I. and Strawderman, Myla H. and Lin, Xi Hong and Kish, Katherine E. and McLaughlin, Patrick W. and Sandler, Howard M.},
    number = {5},
    pages = {941--7},
    volume = {38},
    doi = {10.1016/S0360-3016(97)00082-5},
    issn = {03603016}
}

@article{Taylor2013Real-TimeModels,
    title = {{Real-Time Individual Predictions of Prostate Cancer Recurrence Using Joint Models}},
    year = {2013},
    journal = {Biometrics},
    author = {Taylor, Jeremy M.G. and Park, Yongseok and Ankerst, Donna P. and Proust-Lima, Cecile and Williams, Scott and Kestin, Larry and Bae, Kyoungwha and Pickles, Tom and Sandler, Howard},
    number = {1},
    pages = {206--213},
    volume = {69},
    doi = {10.1111/j.1541-0420.2012.01823.x},
    issn = {15410420}
}

@article{Joolharzadeh2023RecentCardio-Oncology,
    title = {{Recent Advances in Serum Biomarkers for Risk Stratification and Patient Management in Cardio-Oncology}},
    year = {2023},
    journal = {Current Cardiology Reports},
    author = {Joolharzadeh, Pouya and Rodriguez, Mario and Zaghlol, Raja and Pedersen, Lauren N. and Jimenez, Jesus and Bergom, Carmen and Mitchell, Joshua D.},
    number = {3},
    month = {3},
    pages = {133--146},
    volume = {25},
    publisher = {Springer},
    doi = {10.1007/s11886-022-01834-x},
    issn = {15343170},
    pmid = {36790618}
}

@article{Piatek2020RisingCancer,
    title = {{Rising serum CA-125 levels within the normal range is strongly associated recurrence risk and survival of ovarian cancer}},
    year = {2020},
    journal = {Journal of Ovarian Research},
    author = {Piatek, Szymon and Panek, Grzegorz and Lewandowski, Zbigniew and Bidzinski, Mariusz and Piatek, Dominika and Kosinski, Przemyslaw and Wielgos, Miroslaw},
    number = {1},
    pages = {102},
    volume = {13},
    doi = {10.1186/s13048-020-00681-0},
    issn = {17572215}
}

@article{Ye2008SemiparametricApproach,
    title = {{Semiparametric Modeling of Longitudinal Measurements and Time-to-Event Data–A Two-Stage Regression Calibration Approach}},
    year = {2008},
    journal = {Biometrics},
    author = {Ye, Wen and Lin, Xihong and Taylor, Jeremy M.G.},
    number = {4},
    pages = {1238–1246},
    volume = {64},
    doi = {10.1111/j.1541-0420.2007.00983.x},
    issn = {15410420}
}

@article{Wu2023Visit-to-visitAnalysis,
    title = {{Visit-to-visit blood pressure variability and the risk of cardiovascular disease: a prospective cohort analysis}},
    year = {2023},
    journal = {Hypertension Research},
    author = {Wu, Shouling and Tian, Xue and Xu, Qin and Zhang, Yijun and Zhang, Xiaoli and Wang, Penglian and Chen, Shuohua and Wang, Anxin},
    number = {12},
    pages = {2622--2634},
    volume = {46},
    doi = {10.1038/s41440-023-01388-7},
    issn = {13484214}
}

@Manual{JMbayes2,
  title        = {JMbayes2: Extended Joint Models for Longitudinal and Time-to-Event Data},
  author       = {Dimitris Rizopoulos and Pedro {Miranda Afonso} and Grigorios Papageorgiou},
  year         = {2025},
  note         = {R package version 0.5-95, https://github.com/drizopoulos/jmbayes2},
  url          = {https://github.com/drizopoulos/jmbayes2}
}
\end{document}